# Application of Advanced Techniques for Metals Identification and Characterisation

David N Githinji

Department of Manufacturing, Industrial & Textile Engineering, Moi University, P.O Box 3900-30100, Eldoret, Kenya

**Abstract**

The appraisal of metallic materials requires application of advanced characterisation techniques. In this paper, the use has been made of Scanning Electron Microscopy, Transmission Electron Microscopy, X-ray diffraction (XRD) and Energy Dispersive X-ray Spectroscopy techniques to characterise morphological, structural and chemical composition of 316-stainless steel and unknown brass sample. The determined lattice parameter of unknown brass sample was 3.6812Å and consisted of 63% Cu and 37% Zn by weight. The XRD analysis indicated a faced-centred-cubic crystal structure and the sample concluded to be an alpha brass. The measured dislocation density in 316-stainless steel increased with increasing plastic strain and the dislocation structures varied from relatively uniform distribution at low strains to cell-like structures at high strains. The spread of X-ray diffraction peak related linearly with the dislocation content of 316-stainless steel.

**Keywords**: Characterisation, Crystal Structure Dislocation Structure, Diffraction Peak, Stainless Steel,

## 1 Introduction

The appraisal of metallic materials is often necessary in order to establish their physical, chemical and/or mechanical states. There are several techniques available for characterising metals. Microstructural examination at spatial resolutions of a few nm is achieved under Scanning Electron Microscopy (SEM) which uses a beam of electrons of much shorter wavelength than that of the visible light. The SEM uses secondary electrons produced at the beam's interaction volume for topographical imaging (Brundle et al. 1992). Transmission Electron Microscopy (TEM) allows microstructural characterisation at higher resolution than those attainable with SEM. It works by directing a focused beam of electrons onto a thin (<200 nm) sample. The transmitted, elastically and inelastically scattered electrons are directed onto a detector through a series of electromagnetic lenses, to provide both imaging and diffraction information. In the diffraction contrast mode, structural defects can be imaged due to scattering of the incident electron wave. The transmitted and diffracted beams form the bright field and the dark field images, respectively. In the diffraction mode, an electron diffraction pattern from a selected area illuminated by the electron beam is projected onto a detector enabling crystal structure analysis(Williams D.B. and Carter C.B. 1996).Energy Dispersive X-ray Spectroscopy (EDX) is an analytical tool for near surface chemical characterisation of materials. It works on the principle that each atom produces a characteristic X-ray photon on interacting with a high energy beam of electrons. By analysing the emitted X-rays, a qualitative and quantitative chemical analysis can be obtained (Brundle et al. 1992). X-ray diffraction (XRD) is a common technique for the study of crystal structures, phase compositions, stresses, crystal orientations and atomic spacing. It is based on constructive interference of monochromatic X-rays and a crystalline sample.

The metallic material appraised in the current study is brass and 316-stainless steel. Brass is a substitution alloy of zinc and copper (Ohring 1995). The percentage of zinc in the alloy can be varied to give a range of brasses such as alpha, alpha-beta, beta and beta-gamma brass. The alpha brasses contain up to 37% zinc and they have faced-centred-cubic crystal structure. They are ductile, electrically conductive and easily cold worked.  The beta brass contains 50% zinc and has a body-centred-cubic crystal structure. The microstructural details of brass are revealed through grinding and polishing of surface to produce smooth and mirror-like finish which is then chemically etched.

Type 316-stainless steels are iron-chromium-nickel alloys that are characterized by a face-centred cubic (FCC) crystal structure at room temperature. They exhibit good high-temperature strength, corrosion and oxidation resistance (Davis 1997; Marshall 1984) and so they are widely used in industry e.g. in power generating plants. The steel contain a number of alloying elements in supersaturated solid-solution in the austenite phase, which tends to precipitate and form second phases such as carbides and intermetallics at elevated temperatures. The precipitation processes are mainly controlled by solute atom diffusion and so their rates depend on the aging temperature and time.

The plastic deformation of 316-stainless steel occurs at ambient temperature by dislocation glide. The dislocation content of this material following plastic deformation can be assessed from the TEM measurements. Equally, the profile of diffraction peaks from the XRD measurements can be used to surmise the dislocation content of materials. However, the shape of diffraction peak depends on several factors: instrumental broadening, compositional heterogeneity, elastic strain heterogeneity and on mean size of the coherently diffracting domains (Fitzpatrick and Lodini 2003). The variation of diffraction peak full width at half maximum (FWHM) as a function of plastic strain in copper foil under in-situ loading has been explored by (Tang 2007) and showed a clear increase in peak-width with plastic deformation. Nevertheless, the information about the effect of precipitation on variation of diffraction peak FWHM with plastic strain is rather scarce and the current paper aims to investigate this. The paper also aims at studying the application of various advanced characterisation techniques for metallic material identification and characterisation. It focuses on the appraisal of brass morphology, structural and chemical composition while it uses 316-stainless steel for evaluating dislocation structures and how they vary as a function of strain.

## 2 Materials and Methods

### 2.1 Materials

The materials studied in this work included 316-stainless steel and a brass sample of unknown composition and crystal structure. The main elemental composition of the steel based on Optical Emission Spectroscopy measurements is given in Table 1. Tensile specimens of 6 mm diameter and 30 mm gauge length were extracted from the steel and deformed uniaxially on an Instron-8862 machine by 5.1%, 9.6%, and 18.4% true strain at room temperature.

Table 1: Main elemental composition of 316H stainless steel studied

| C | Si | Mn | P | S | Cr | Ni | Mo | N |
|---|---|---|---|---|---|---|---|---|
| 0.07 | 0.42 | 1.00 | 0.03 | 0.02 | 17.82 | 11.81 | 2.33 | 0.1 |

### 2.2 Methods

#### 2.2.1 X-rays Diffraction (XRD) Measurements

The brass and steel samples were prepared for analysis by mechanical polishing to obtain a smooth and flat surface. The samples were then mounted on the XRD machine (theta/2theta) and scanned at slow rate using X-rays of 1.5406Å wavelengths. The samples were tilted at 0° and 90° during the scanning process and diffraction patterns taken for analysis. The X-ray diffraction peaks obtained from the stainless steel samples were analysed using X-ray stress Analyser (XSTRESS 3000) software from which the Full Width at Half Maximum (FWHM) values were determined for strained and unstrained samples.

#### 2.2.2 Scanning Electron Microscopy (SEM) Measurements

The brass and steel samples were prepared for measurement by grinding sequentially with SiC papers followed by mechanical polishing with diamond paste down to a 1μm. The final preparation stage was etching which enhanced visualisation of microstructures. A fractured brass surface was also appraised. The prepared samples were mounted on the SEM machine and secondary electron imaging performed using an acceleration voltage of 20 kV. The SEM images were captured at different magnifications for further analysis.

#### 2.2.3 Energy Dispersive X-ray (EDX) Measurements

The brass sample preparation was similar to that adopted for SEM measurement. The sample was mounted on the EDX machine and its net X-ray counts recorded. This was also repeated for pure copper, pure zinc and standard brass samples.

#### 2.2.4 Transmission Electron Microscopy (TEM) Measurements

The sample preparation for TEM measurement consisted of machining 0.3 mm thick slices from all deformed steel specimens and punching them to produced 3 mm diameter discs. The discs were then ground sequentially using 40 μm, 15 μm and 5 μm grit SiC paper until a thickness of about 100 μm was achieved. The final preparation entailed electropolishing the discs to perforation using a potential difference of 20V at -50±5 °C and an electrolyte consisting of 5% perchloric acid in methanol. The prepared samples were appraised using a JEOL JEM 2100 TEM with a $LaB_6$ emitter and a maximum lateral resolution of 0.25 nm. It was operated at 200 keV with images acquired using a Gatan Orius SC1000 digital camera mounted in-line with the TEM optics. Measurements were done at different location of the samples with different imaging vectors. Imaging was carried out in the bright field mode and the corresponding diffraction pattern recorded using a selected area diffraction mode. The diffraction contrast TEM images were used for calculating dislocation densities following procedures given in (Ham 1961).

## 3  Results and Discussion

### 3.1  Morphological Assessment

The SEM images provided the sample's morphological information such as the grain sizes, grain orientations, inclusions and the grain boundaries as shown in Fig. 1(a). From the SEM images, the details about mechanical history of a material can be deduced. As seen in Fig. 1(a) it is possible that the brass sample was annealed before imaging owing to the presence of twins and fine grains in the microstructure. Annealing twins are known to develop in materials with FCC crystal structure (Dieter 1988). The stretched grains shown in Fig. 1(b) are an indication of tensile loading on the material before imaging. The details of brass fracturing pattern which are characteristic of a ductile failure are shown in Fig. 1(c) with the bright and dark areas representing raised and deep points on the sample's surface, respectively. The microstructure of 316-stainless steel exhibited an extensive intra- and inter-granular precipitation as seen in Fig. 1(d). The intergranular precipitate appeared relatively large with high aspect ratio compared to the intragranular precipitates which displayed a circular morphology. The observed morphologies are common in over-aged 316-stainless steel (Hong et al. 2001; Rios and Padilha 2005; Weiss and Stickler 1972).

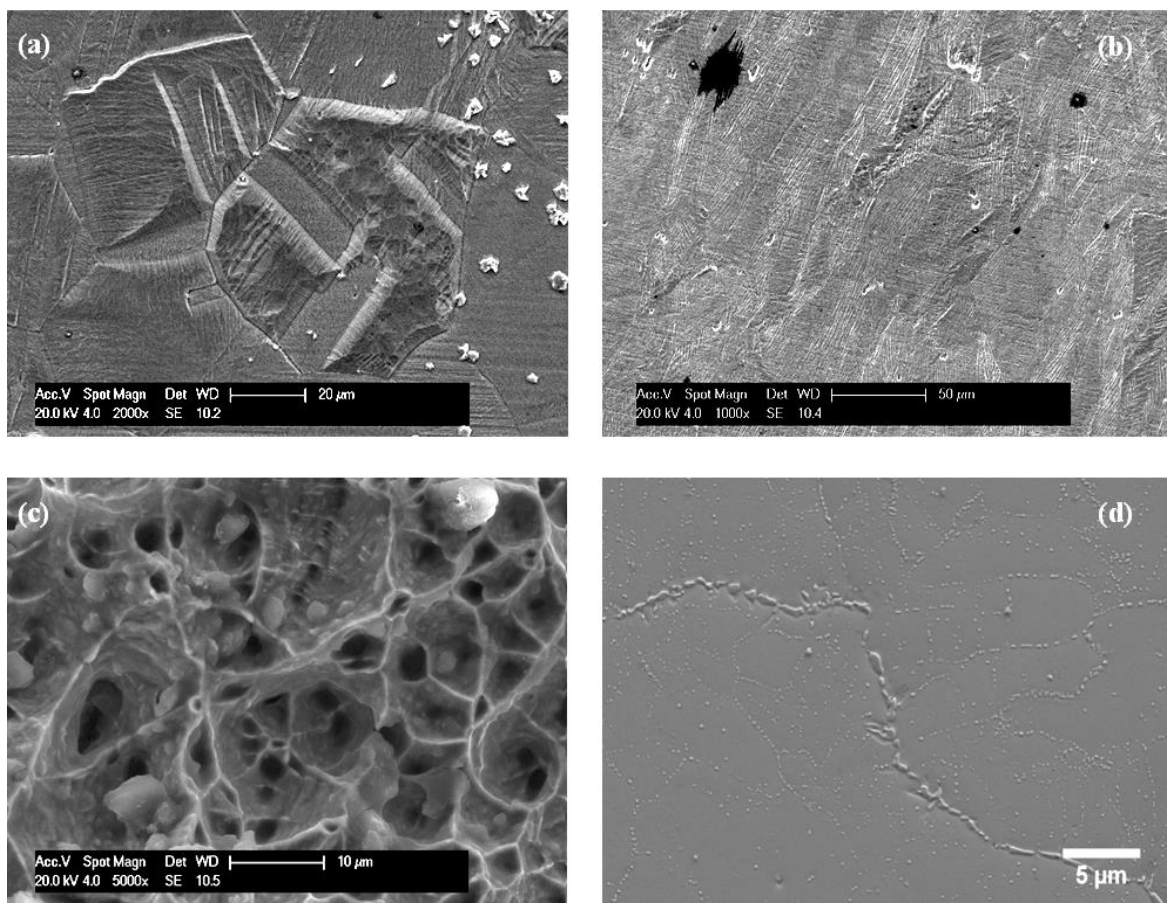

Fig. 1. SEM images of: (a) polished and etched brass sample, (b) fractured annealed brass surface showing stretching of grains, (c) fractured brass surface indicating ductile failure, and (d) 316-stainless steel showing inter- and intra-granular precipitates.

### 3.2 Structural Assessment

The XRD technique was used to obtained information about the crystal structure of the unknown brass sample. The slow scan diffraction pattern shown in Fig. 2, exhibited a repeat of two peaks in succession followed by a single peak. This kind of diffraction pattern is a characteristic of FCC crystal structure (Suryanarayana and Norton 1998). Consequently, it may be concluded that the unknown brass sample had this kind of crystal structure and was of alpha brass type.

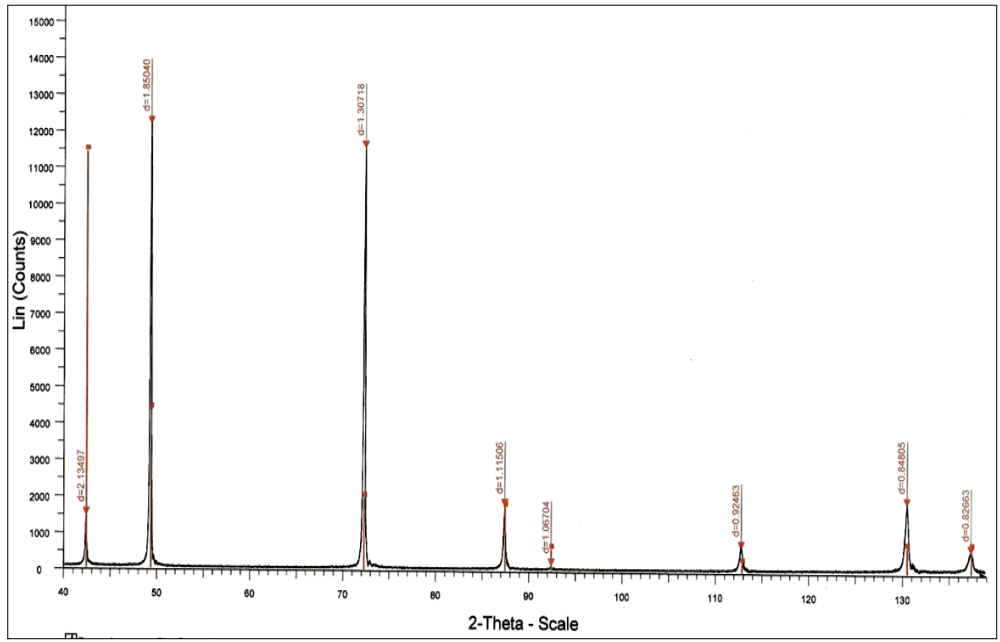

Fig. 2. XRD diffraction patterns for annealed brass sample at slow scanning rate showing diffraction angles against diffracted beam intensity.

Table 2 shows the determined lattice parameter and indices of planes producing each peak in the diffraction patterns in Fig. 2. By applying the lattice parameter, the atomic concentration of zinc in the alloy was estimated as 30% using Nelson-Riley analysis (Waseda et al. 2011). However, this technique is known to be relatively inaccurate when used in sample's composition analysis.

Table 2: The unknown brass sample lattice parameter and diffracting (*hkl*) planes

| Peak | 2θ | $Sin^2θ$ | $Sin^2θ/ 0.0142$ | $h^2+k^2+l^2$ | (hkl) | Lattice parameter |
|---|---|---|---|---|---|---|
| 1 | 0.2361 | 0.0139 | 0.98 | 3 | (111) | 3.6812 |
| 2 | 0.2750 | 0.0188 | 1.32 | 4 | (200) | |
| 3 | 0.4000 | 0.0395 | 2.78 | 8 | (220) | |
| 4 | 0.4861 | 0.0579 | 4.08 | 11 | (311) | |
| 5 | 0.5139 | 0.0646 | 4.55 | 12 | (222) | |
| 6 | 0.6278 | 0.0953 | 6.71 | 16 | (400) | |

The dislocation structures in the deformed steel samples were assessed through TEM imaging. As seen in Fig. 3, dislocation structures in the specimen with no deformation consisted of individual dislocations which were fairly randomly distributed. However, as the strain was increased there was evidence of diffuse cell-wall structures and localized dislocation tangles which became more pronounced at higher strain. The increase in the dislocation content with strain may be attributed to dislocation generation, multiplication and interactions (Dieter 1988).

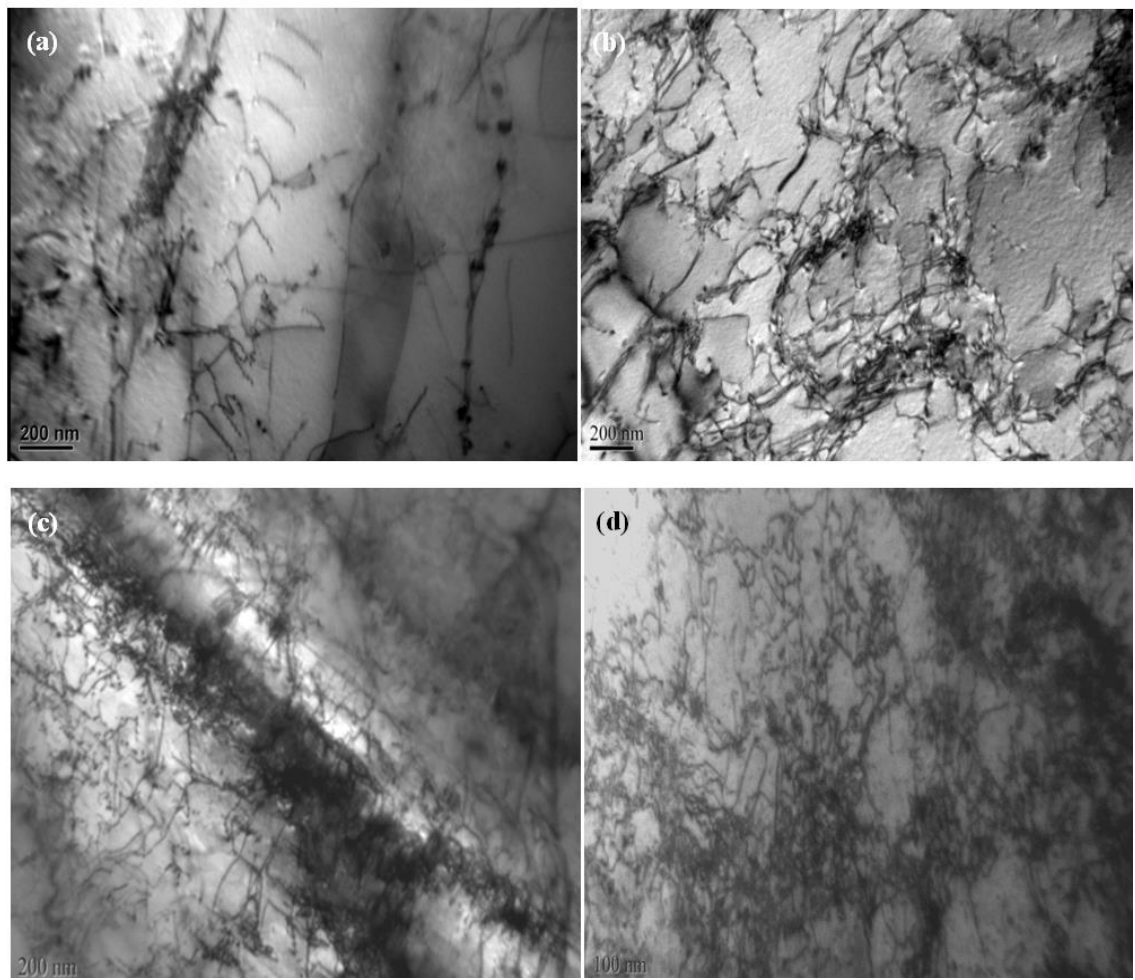

Fig. 3. TEM BF micrographs showing dislocations in steels deformed by: (a) 0, (b) 0.051, (c) 0.096 and (d) 0.184 true strain.

The calculated average dislocation densities for deformed 316-stainless steels are given in Table 3. The dislocation density increased almost linearly with the increase in plastic deformation as seen in Fig. 4. Consequently, the measured dislocation density can provide information about the mechanical history of a metal provided the starting dislocation density is known.

Table 3: Average dislocation densities $\rho$ (m$^{-2}$) in 316-stainless steel based on TEM measurement.

| Sample ID | True strain | $\rho$ by TEM |
|---|---|---|
| A | 0 | 8.3E+13 |
| B | 0.051 | 1.9E+14 |
| C | 0.096 | 3.1E+14 |
| D | 0.184 | 5.1E+14 |

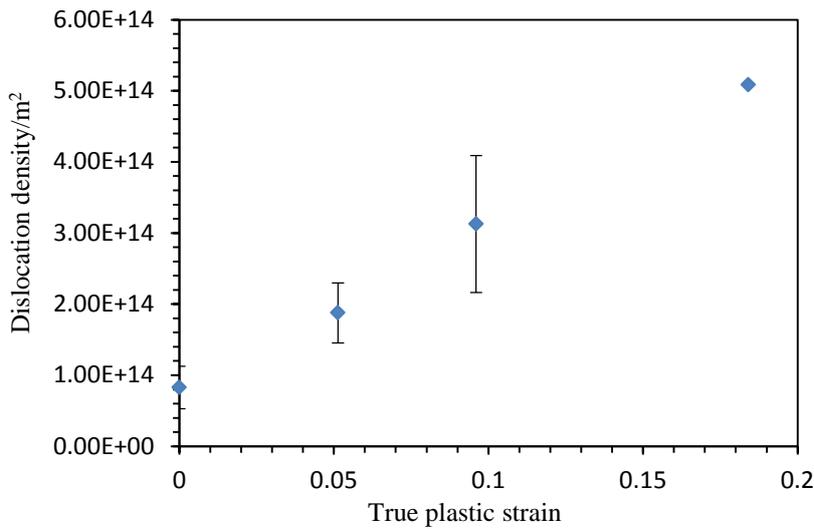

Fig. 4. Stainless steel measured dislocation density versus true plastic strain.

The relationship between the FWHM and plastic strain is shown in Fig. 5(a) and that between FWHM and dislocation density in Fig. 5(b). It is evident that the spread in X-ray diffraction peaks relates in a linear manner to the density dislocations in the microstructure. However, this spread may also be attributed to other factors (Fitzpatrick and Lodini 2003) and it is therefore erroneous to attribute the spread entirely to the plastic deformation in the material. However, the FWHM measurements can provide a qualitative assessment of the dislocation content of similar metallic material.

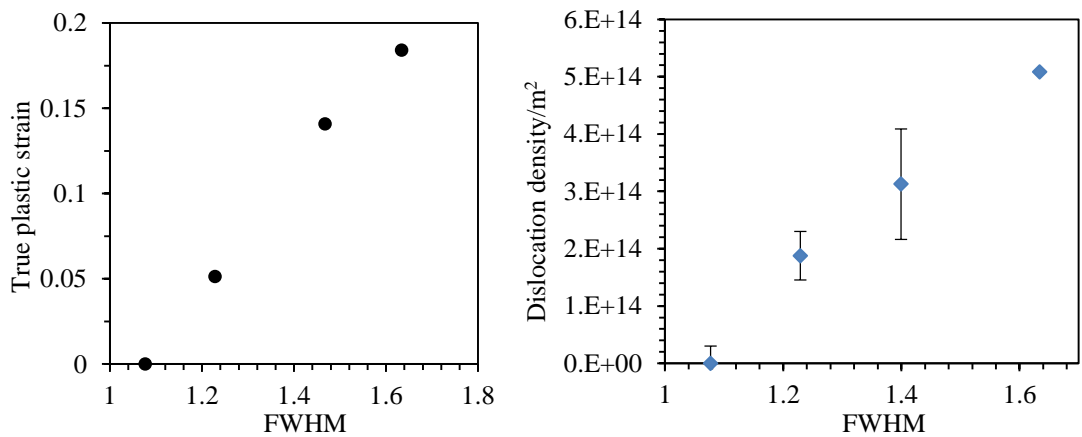

Fig. 5. Variation of FWHM at 0 ° tilts with: (a) plastic strain, (b) dislocation density in 316-stainless steels.

### 3.3 Chemical Composition Measurement

The EDX technique was used to appraise the chemical composition of the unknown brass sample. Table 4 gives the X-rays counts obtained from the EDX spectra for the unknown brass, pure copper, pure zinc and from a standard brass sample of 70 wt% Cu: 30 wt% Zn. From the composition of the standard brass sample, the determined X-ray count values were used to compute the composition of Cu and Zn in the unknown brass sample. The sample was found to have 63.2% Cu and 36.8% Zn by weight after ZAF correction. The main challenge encountered during this measurement was poor spectra resolution.

Table 4: Mean X-ray counts for pure copper, pure zinc, standard brass and unknown brass sample

| Sample | Cu Ka window Net X-ray counts | Zn Ka window Net X-ray counts |
|---|---|---|
| Pure Cu standard | 6724 | |
| Pure Zn standard | 6727 | |
| 70 Cu:30 Zn brass standard | 6673 | 1803 |
| Unknown brass specimen | 5990 | 2263 |

## 4 Conclusion

Application of advanced metal characterisation techniques have been explored in the current work with specific reference to brass and 316-stainless steel. The following conclusions were drawn from the work:

- A combination of analytical techniques is required for complete characterisation of metals: SEM imaging provide morphological information while EDX give information concerning elemental composition of the metal. The crystal structure and the distortion in the crystal lattice are studied using XRD while dislocation structures are evaluated using TEM imaging.
- Appraisal of unknown brass sample through application of advanced characterisation techniques determined its lattice parameter as 3.6812, its crystal structure as FCC and its composition as 63% copper and 37% zinc by weight.
- The measured dislocation density in 316-stainless steel increases with increasing plastic strain. This is thought to arise from generation, multiplication and interaction of dislocations as deformation in the material increases.
- The spread of X-ray diffraction peak relates linearly to the dislocation content of 316-stainless steel, thus indicating the potential of FWHM measurement for assessing the degree of plastic strain in similar material.